\documentclass[lineno]{jfm}

\usepackage{graphicx}
\graphicspath{{figs_1/}}
\usepackage{ulem}
\usepackage{newtxtext}
\usepackage{newtxmath}
\usepackage{natbib}
\usepackage{multirow}
\usepackage{hyperref}
\hypersetup{
    colorlinks = true,
    urlcolor   = blue,
    citecolor  = black,
}

\newcommand{\RomanNumeralCaps}[1]
\linenumbers

\newcommand\Ray{\mbox{\textit{Ra}}}
\newcommand\Nuy{\mbox{\textit{Nu}}}

\title{Physical mechanism of the convective heat flux increasing in case of mixed boundary conditions}

\author{Andrei Sukhanovskii\aff{1}
\corresp{\email{san@icmm.ru}} \and Andrei Vasiliev\aff{1}}
\affiliation{\aff{1}Institute of Continuous Media Mechanics,  Korolyov 1, Perm, 614013, Russia}

\begin{document}
\maketitle

\begin{abstract}
A series of numerical simulations of Rayleigh-B{\'e}nard convection in a cubic cavity are conducted in order to examine the structure of the thermal boundary layer in case of mixed boundary conditions. The main goal of the study is the physical mechanism which provides increasing of heat flux with spatial frequency of conducting-adiabatic pattern. Different spatial configuration of conducting plates, including the fractal one, are considered for Rayleigh numbers from $\Ray=10^{7}$ to $\Ray=2.0\times 10^{9}$. We have shown that the temperature boundary layer in case of mixed boundary conditions at the bottom is strongly non-uniform. This non-homogeneity is a result of several factors such as conducting-adiabatic pattern, large-scale circulation and small-scale motions over conducting plates. The thickness of the thermal boundary layer strongly depends on the size of the conducting plates and can be substantially smaller than for a classical Rayleigh-B{\'e}nard convection. This effect increases the heat flux with decreasing the size of hot plates, which corresponds to the increasing of spatial frequency of conducting-adiabatic pattern. 
\end{abstract}

%\begin{keywords}
%Rayleigh-B{\'e}nard convection, mixed boundary conditions, heat transfer
%\end{keywords}

%{\bf 76F35}

\section{Introduction}
\label{sec:intro}

Thermal convection is the main source of the motion in atmosphere and oceans and understanding of its fundamental aspects is necessary for solution of multiple problems emerging in geophysical and technological systems. Scientific efforts are mainly focused on the thermal convection in the enclosure heated from below and cooled from above ~\citep{AhlersRevModPhys2009, ChillaEPJE2012, XiaTAML2013}, which is known as Rayleigh-B{\'e}nard convection (RBC). Classical RBC is assumed that the temperatures on the upper and lower boundary are constant and have uniform distribution. A wide spectre of applied problems leads to the studies of more complex cases that include different cell geometries, boundary conditions ~\citep{DasIHMT2017, EvgrafovaIHMT2019, NandukumarPoF2019, PandeyIHMT2019} and presence of both horizontal and vertical temperature gradients ~\citep{ZiminFD1982, FabregatIHMT2020}.

The specific case when thermal boundary conditions are inhomogeneous is of special interest. Geophysical studies of the role of inhomogeneous temperature boundary conditions are concentrated on the mantle convection, because continental plates are less heat conductive in comparison with the oceanic ones ~\citep{JaupartElsevier2015} and can be treated as thermal insulators. The feedback between the convective flows and moving lithospheric plates leads to the complex dynamic of the plate tectonics ~\citep{Whitehead2015GRL}. The influence of one or more insulating blocks on the heat transfer efficiency was studied in ~\citep{CooperGRL2013}. Significant differences in heat transfer for various configurations of insulated regions were observed when the total area of the blocks exceeded 50\%. For a single block, the Nusselt number was always lower than for the several small blocks. The experiments of~\citep{WangJFM2017} showed that the overall heat transfer efficiency decreases with increasing area of the heat-insulating plate at the upper boundary and weakly depends on the spatial distribution of the insulating plates. The role of spatial frequency of insulator distribution was revealed in ~\citep{RipesiJFM2014, DennisJFM2018}.  It was observed that heat transfer efficiency increases with adiabatic pattern frequency and for the scale of the insulators comparable to the thickness of the temperature boundary layer, the Nusselt number is very close to the one in case of classical Rayleigh-B{\'e}nard convection. Physical interpretation of this effect was proposed in ~\citep{DennisJFM2018} as follows. The thermal boundary layer masks actual boundary including insulated regions and presents a new effective boundary to the bulk flow. The horizontal heat flux from the conducting regions in the thermal boundary layer increases with a pattern frequency. 

In the present paper we focus our attention on the structure of the thermal boundary layer and physical mechanism of variation of the heat flux with adiabatic pattern structure and frequency. We have found that key role plays non-homogeneous thermal boundary layer over conducting regions. The thickness of the thermal boundary layer is significantly lower in the peripheral parts of conducting plates which remarkably increases local heat fluxes and as a result provides substantial growth of Nusselt number.

\section{Numerical simulation}
\label{sec:numerical}

\begin{figure}
\center{\includegraphics[width=0.50\linewidth]{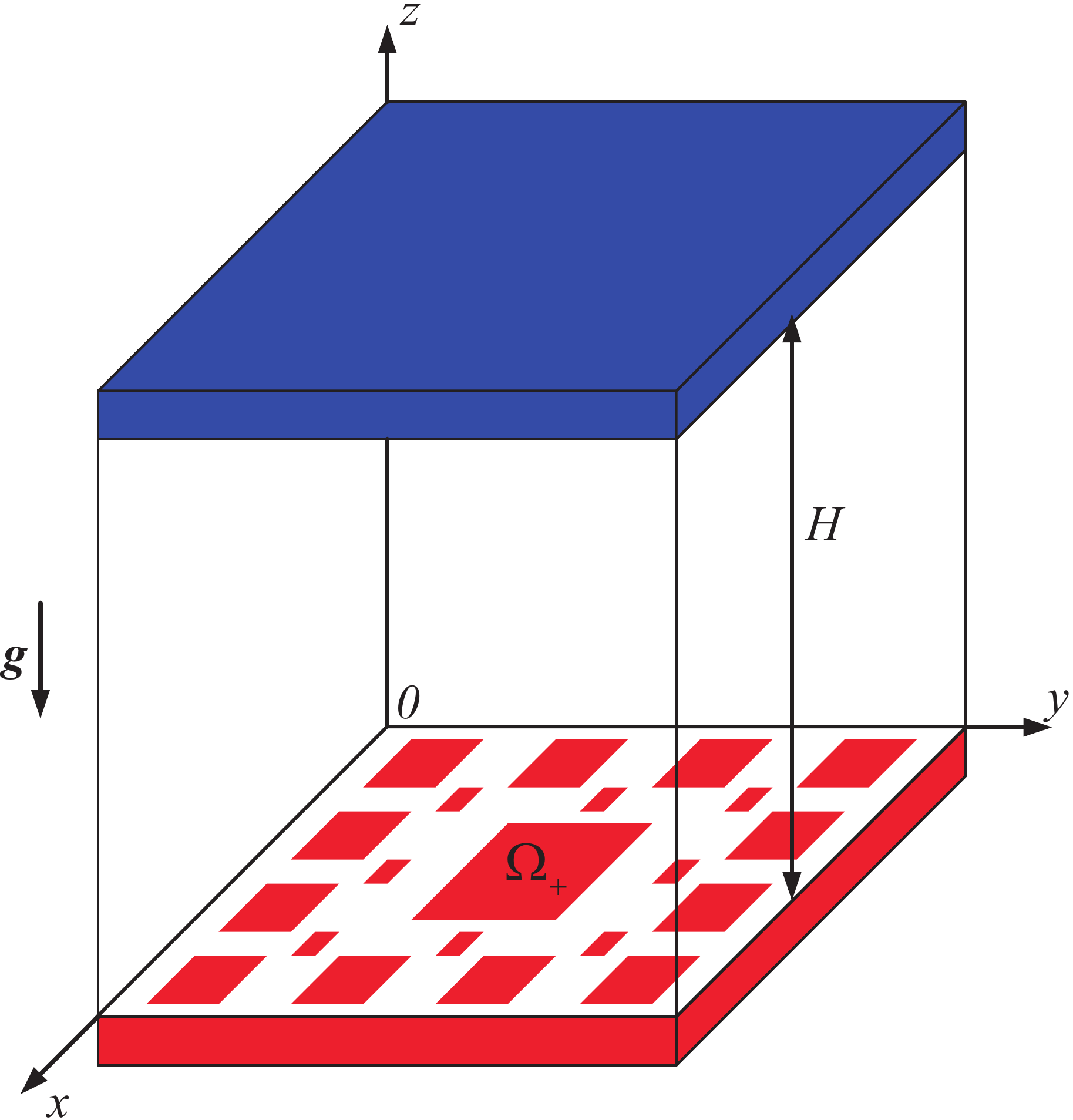}}
\caption{Sketch of the computational domain.}
\label{fig:Schema}
\end{figure}

We study the features of heat transfer of an incompressible, viscous fluid in a cubic cavity of size $H$ under homogeneous and inhomogeneous thermal boundary conditions (see figure~\ref{fig:Schema}).
\begin{figure}
\vspace{0.35cm}(\textit{a})\hspace{4.1cm}(\textit{b})\hspace{4.1cm}(\textit{c})\\\vspace{-0.2cm}\\
\hspace{0.3cm}\includegraphics[width=0.30\linewidth]{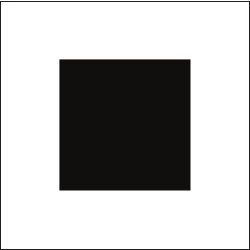}
\hspace{0.3cm}\includegraphics[width=0.30\linewidth]{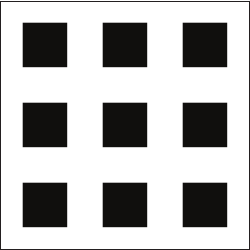}
\hspace{0.3cm}\includegraphics[width=0.30\linewidth]{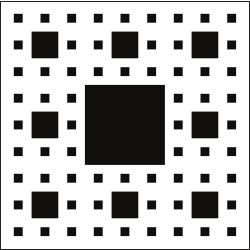}
\caption{Three configurations of the heating: (\textit{a}) localized heating in the center, (\textit{b}) nine heaters of the same size, equidistant from each other and (\textit{c}) three steps of Sierpinski carpet. Black regions correspond to a constant temperature, and white regions to adiabatic conditions.}
\label{fig:NonHom}
\end{figure} 
For homogeneous heating isothermal conditions at the horizontal plates are used. Inhomogeneous heating is applied only at the lower boundary by using mixed boundary conditions.  The lower boundary included heating regions $\Omega_{+}$ that are maintained at constant temperature $T_{b}$ while the remaining part of the bottom was thermally insulated. Figure~\ref{fig:NonHom} shows three configurations of the $\Omega_{+}$ distribution: (I) - localized heating in the center; (II) - nine heaters of the same size, equidistant from each other and (III) - a combination of $\Omega_{+}$ of three sizes with an inhomogeneous spatial distribution. Note that the last configuration is a pattern of Sierpinski carpet of third order. All configuration have the same heating area, which is $30\pm0.5 \%$ of the bottom.

The problem of thermal convection is numerically investigated within the Boussinesq approximation at a low Rayleigh number $\Ray=10^{7} $ without using turbulence models. It is described by the following equations in the Cartesian coordinates:
\begin{gather}
\frac{\partial u_{i}}{\partial x_{i}}= 0,  \label{eqn:incom}\\
\frac{\partial u_{i}}{\partial t} + \frac{\partial}{\partial x_{j}}\left(u_{i}u_{j}\right) = -\frac{\partial}{\partial x_{i}}\left(\frac{p}{\rho_{0}}\right)  + \nu_{0}\frac{\partial^{2} u_{i}}{\partial x_{j} \partial x_{j}} + \left(\frac{\rho}{\rho_{0}}\right)g\delta_{i3}, \label{eqn:momen}\\
\frac{\partial T}{\partial t} + \frac{\partial}{\partial x_{j}}\left(Tu_{j}\right) = \chi_{0}\frac{\partial^{2} T}{\partial x_{j} \partial x_{j}}, \label{eqn:energy}\\
\frac{\rho}{\rho_{0}} = 1 - \beta (T - T_{0}) \label{eqn:state},
\end{gather}
where $u_{i}$ is the $i$th component of the velocity field, $p$ is the pressure, $T$ is the temperature, $\rho$ is the density, $\beta$ is the thermal expansion coefficient, $g$ is the gravity acceleration, $\delta_{i3}$ is the Kronecker symbol and $\rho_{0}$, $\nu_{0}$, $\chi_{0}$ are the density, kinematic viscosity, thermal diffusivity at the reference temperature $T_{0}$, respectively. For the non-dimensionalization, we have used the height of the cube $H$ as the length scale, $U_{f}=\sqrt{\beta g \Delta T H}$ (free-fall velocity) as the velocity scale, and $\theta=(T-T_{t})/\Delta T$ as the temperature scale.

In fully developed turbulence, resolution of all relevant flow scales using direct numerical simulation (DNS) requires large computational resources. One of the alternative approaches is the use of the large eddy simulation (LES) method, which allows to avoid explicit resolution of the Kolmogorov scale $\eta_{k}$. Therefore, in the case of moderate and high Rayleigh numbers $10^{8} < \Ray < 2.0\times 10^{9} $ the LES approach is used. Equations ~\eqref{eqn:incom}--~\eqref{eqn:state} are filtered through application of a low-pass filter whose width is proportional to the cell size $\Delta=(\Delta_{x}\Delta_{y}\Delta_{z})^{1/3}$. The filtered equations contain two unknown terms $\tau_{ij}=\widetilde{u_{i}u_{j}}-\tilde{u}_i\tilde{u}_j$ and $\lambda _ {i}=\widetilde{\theta u_{i}}-\tilde{\theta}\tilde{u}_i$, which are the subgrid-scale (SGS) stress tensor and the subgrid-scale heat flux, respectively. Here, ($\sim$) denotes the filtering operation, $\tilde{u}_i$ is the resolved velocity and $\tilde{\theta}$ is the resolved temperature. These terms provide energy dissipation at the length scale approximately equal to the filter width. To close the equations we use the Smagorinsky model to parameterize the SGS stress tensor
\begin{equation}
  \tau_{ij}=-2(C_{s}\Delta)^{2}\vert \tilde{S} \vert \tilde{S}_{ij}=-2\nu_{sgs}\tilde{S}_{ij}
  \label{Sgs_st}
\end{equation}
and a simple eddy diffusivity model in which the SGS heat flux is proportional to the resolved temperature gradient
\begin{equation}
  \lambda_{i}=-\kappa_{sgs}\partial \tilde{\theta}/\partial x_{i},
  \label{Sgs_hf}
\end{equation}
where $\tilde{S}_{ij}=1/2\left(\partial \tilde{u}_{i}/\partial x_{j} + \partial \tilde{u}_{j}/\partial x_{i} \right) $ is the strain rate tensor, $\vert \tilde{S} \vert=\sqrt{2\tilde{S}_{ij}\tilde{S}_{ij}}$  is the norm of the strain rate tensor, $C_{s}=0.18$ is the Smagorinsky coefficient, $\nu_{sgs}=(C_{s}\Delta)^{2}\vert \tilde{S} \vert$ is the SGS viscosity and $\kappa_{sgs}$ is the SGS thermal diffusitvity. SGS thermal diffusivity is calculated as
\begin{equation}
  \kappa_{sgs}=\nu_{sgs}/\Pran _{sgs},
  \label{Sgs_Pr}
\end{equation}
where $\Pran _{sgs}$ is the SGS Prandtl number. In our study, we use the constant $\Pran _{sgs}$ approach, which assumes a single constant value for $\Pran _{sgs}=0.9$ in the entire computational domain.

The thermal convection equations for both DNS and LES approaches are solved numerically using the open-source finite volume code OpenFOAM 4.0~\citep{WellerCP1998}. Convective terms and diffusion terms in the equations are approximated by Gauss linear and Gauss linear corrected schemes, respectively. These schemes have the second order of approximation accuracy. The time derivative is approximated by the second order an implicit ``backward'' scheme. The time step is adaptive, the Courant-Friedrichs-Lewy (CFL) number does not exceed 0.4. Pimple algorithm (combination of PISO and SIMPLE) is used for pressure-velocity coupling. The computational grid is a structured mesh. The grid is additionally refined near the horizontal boundaries to resolve the boundary layers. The number of nodes in the temperature and velocity boundary layers  were chosen according to the criterion suggested in~\citep{ShishkinaNJP2010}. A list of all simulations, their grid parameters and Nusselt numbers $\Nuy$ are presented in table ~\ref{tab:1}.

\begin{table}
\begin{center}
\def~{\hphantom{0}}
\begin{tabular}{lcccccc}
Configuration        & $\Ray$              & $N_{x}\times N_{y}\times N_{z}$            & $N_{u}$ & $N_{T}$ & $max(\Delta/ \eta_{k})$  & $\Nuy$ \\[3pt]
\multirow{3}{*}{I}   & $10^{7}$            & \multirow{3}{*}{$140\times 140\times 140$} & 11/5    & 12/3    & 0.31                  & 9.1\\
                     & $10^{8}$            &                                            & 8/8     & 7/4     & 0.63                  & 16.0\\
                     & $1.1\times 10^{9}$  &                                            & 5/11    & 5/11    & 1.50                  & 36.0\\
                     & $1.1\times 10^{9}$  & $272\times 272\times 272$                  & 9/11       & 6/6       & 0.77                           & 32.9\\
\hline
\multirow{3}{*}{II}  & $10^{7}$            & \multirow{3}{*}{$132\times 132\times 132$} & 14/5    & 12/3    & 0.33                  & 10.0\\
                     & $10^{8}$            &                                            & 9/7     & 7/4     & 0.69                  & 18.0\\
                     & $1.1\times 10^{9}$  &                                            & 5/11    & 3/6     & 1.49                  & 35.7\\
                     & $1.1\times 10^{9}$  & $274\times 274\times 274$                  & 9/11    & 6/6     & 0.78                         & 34.2\\
\hline
\multirow{4}{*}{III} & $10^{7}$            & \multirow{3}{*}{$135\times 135\times 135$} & 14/5    & 13/3    & 0.33                  & 10.8\\
                     & $10^{8}$            &                                            & 9/7     & 7/4     & 0.68                  & 19.3\\
                     & $1.1\times 10^{9}$  &                                            & 5/11    & 3/6     & 1.47                  & 37.8\\
                     & $1.1\times 10^{9}$  & $270\times 270\times 270$                  & 8/11    & 6/6     & 0.80                         & 36.6\\
\hline
\multirow{4}{*}{RBC} & $10^{7}$            & \multirow{3}{*}{$135\times 135\times 135$} & 14/3    & 15/3    & 0.50                         & 15.8\\
                     & $10^{8}$            &                                            & 12/7    & 9/4     & 1.06                         & 32.5\\
                     & $1.1\times 10^{9}$  &                                            & 7/11    & 4/6     & 2.37                         & 74.6\\
                     & $1.1\times 10^{9}$  & $270\times 270\times 270$                  & 14/11   & 10/6    & 1.16                         & 67.8\\
\end{tabular}
\caption{Simulation parameters for $\Pran=6.46$. $N_{x}$, $N_{y}$ and $N_{z}$ are number of grid points along $x$, $y$, and $z$ directions; numbers of grid points required to resolve the thermal and the viscous boundary layers are $N_{T}$ and $N_{u}$ respectively (actual resolution/requirement); $max(\Delta/ \eta)$ is maximum local cell size compared to the Kolmogorov length scale $\eta_{k}\approx \Pran^{1/2}/((\Nuy-1)\Ray)^{1/4}$; global Nusselt number $\Nuy=1+\sqrt{\Ray \Pran}\langle u_{z}\theta \rangle_{V,t}$.}
\label{tab:1}
\end{center}
\end{table}

No-slip velocity conditions are applied at all boundaries ($u_{i}=0$ and $\tilde{u}_i=0$). Adiabatic conditions are used for the lateral walls. Thermal boundary conditions for horizontal boundaries can be written as follows:
\begin{equation}
\begin{aligned}
& \theta(x, y, z=0) = 1, & \forall \; x, y \in \Omega_{+}, \\
& \partial_{z}\theta(x, y, z=0) = 0, & \forall \; x, y \notin \Omega_{+}, \\
& \theta(x, y, z=1) = 0, & \forall \; x, y.
\end{aligned}
\end{equation}

Application of asymmetric thermal boundary conditions leads to the disbalance of the heat flux through the lower and upper boundaries.  Therefore, all the results presented below are for the stage of a quasi-steady-state heat balance. The time averaging is performed over at least 200 convective time units.

\section{Results}
\label{sec:results}

\begin{figure}
\vspace{0.35cm}\hspace{0.6cm}(\textit{a})\hspace{4.2cm}(\textit{b})\hspace{4.2cm}(\textit{c})\\\vspace{-0.3cm}\\
  \includegraphics[width=0.33\linewidth]{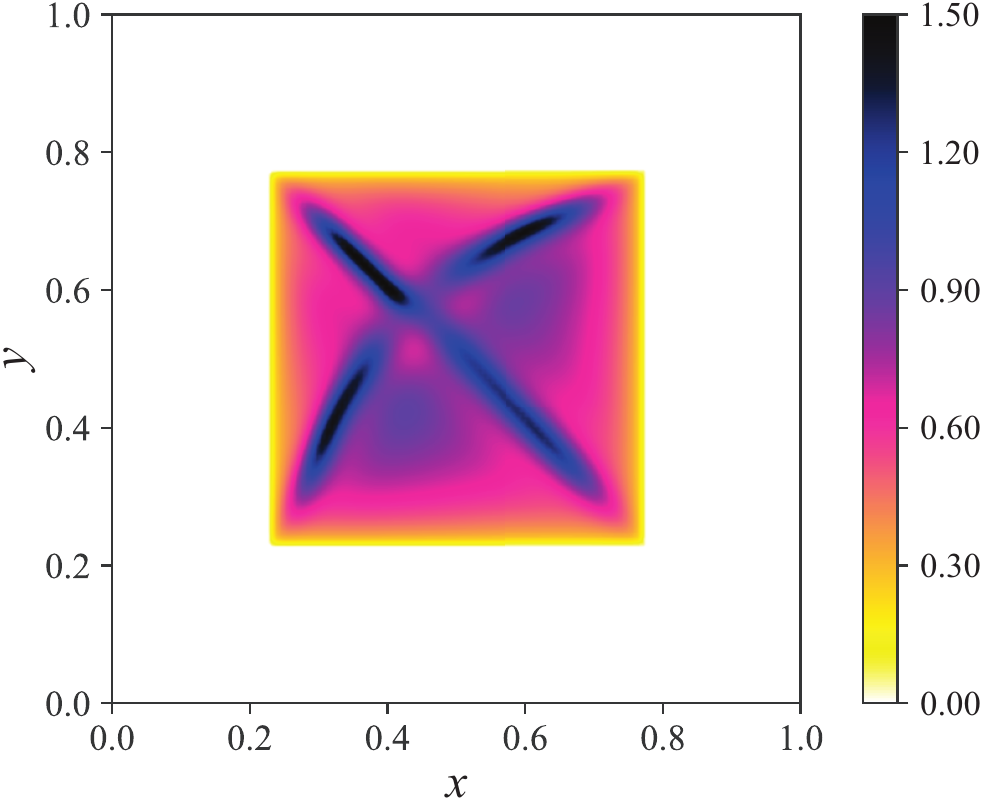}
  \includegraphics[width=0.33\linewidth]{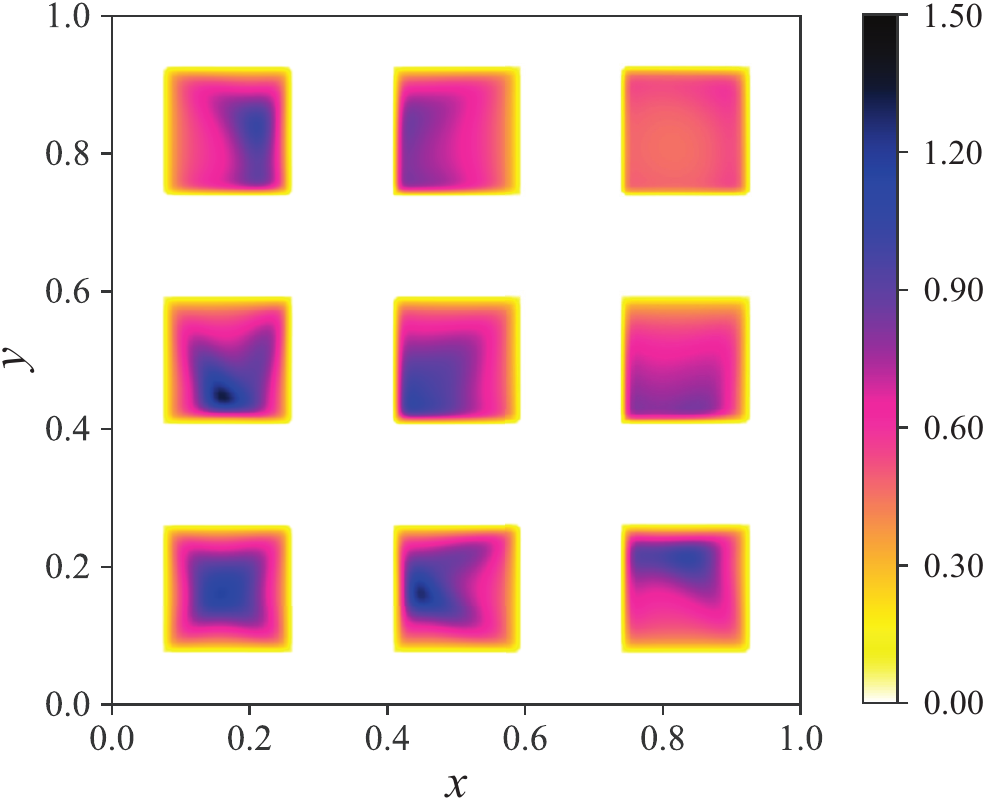}
  \includegraphics[width=0.33\linewidth]{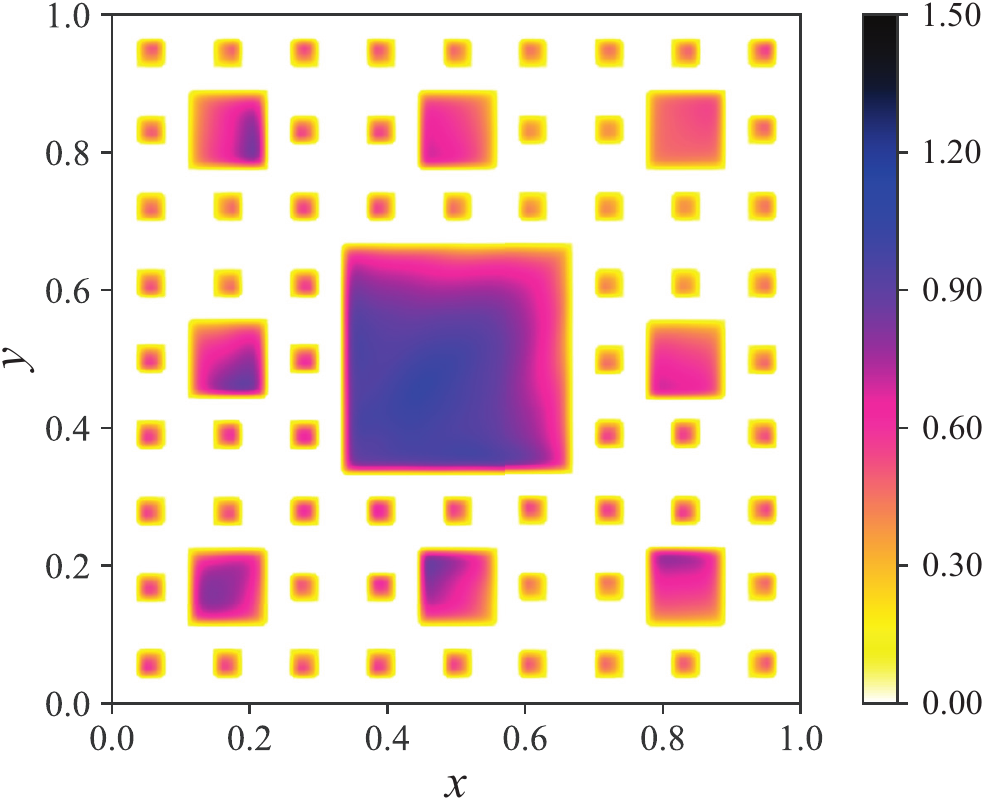}
\vspace{0.0cm}\hspace{0.6cm}(\textit{d})\hspace{4.2cm}(\textit{e})\hspace{4.2cm}(\textit{f})\\\vspace{-0.3cm}\\
  \includegraphics[width=0.33\linewidth]{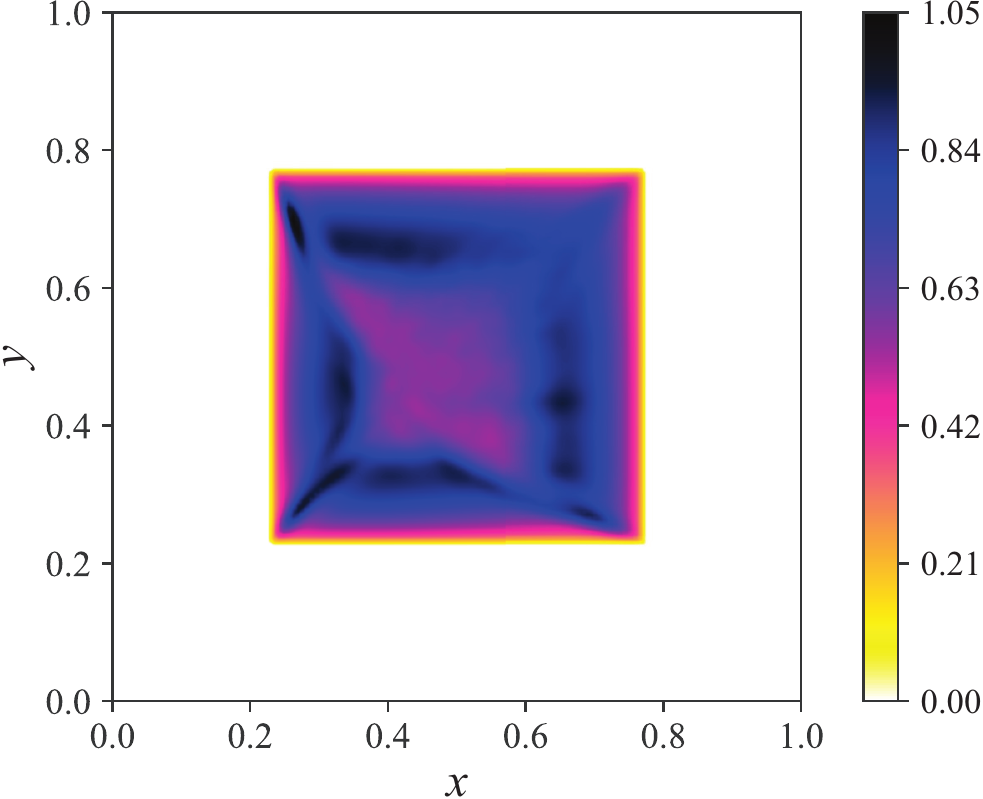}
  \includegraphics[width=0.33\linewidth]{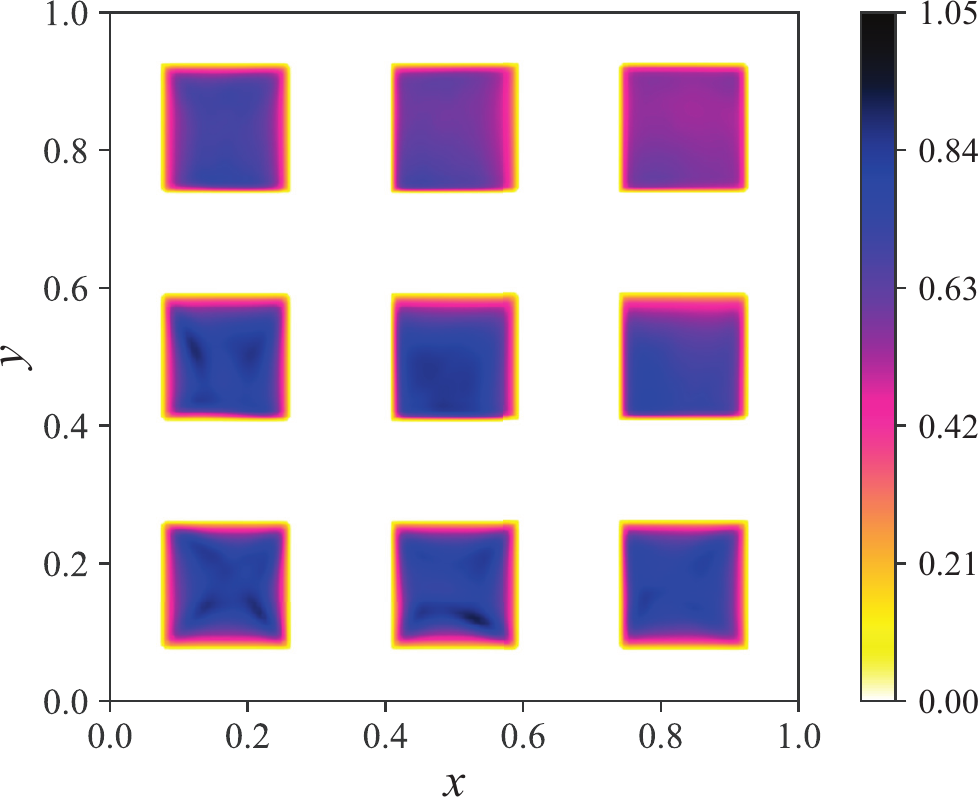}
  \includegraphics[width=0.33\linewidth]{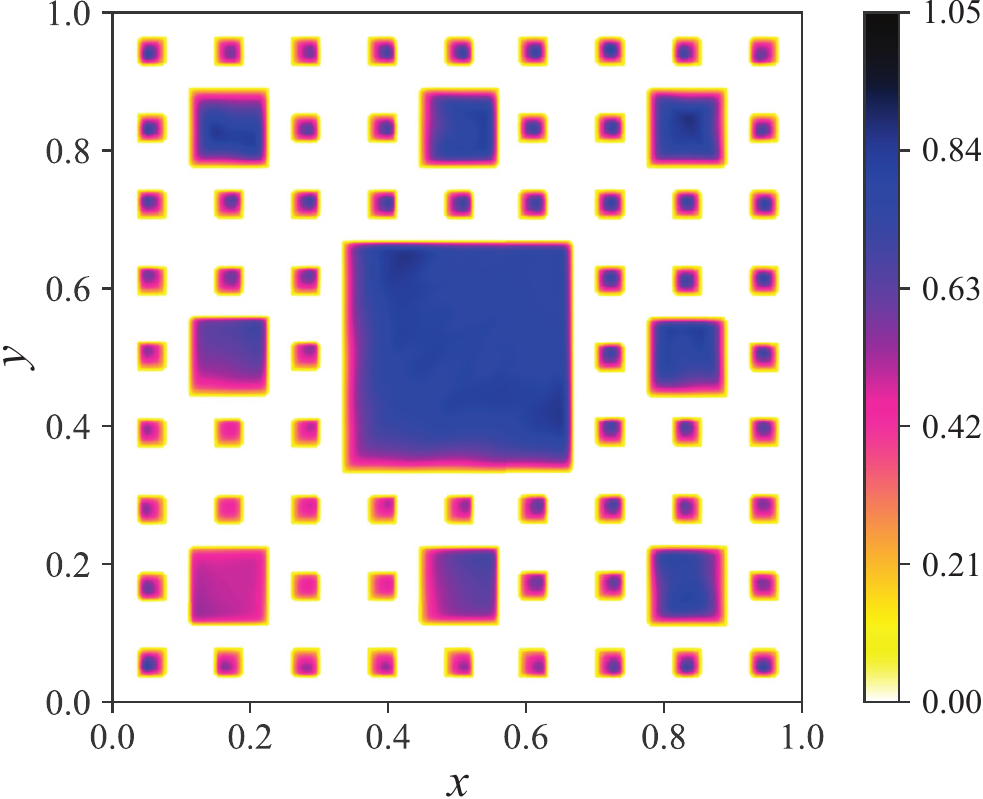}
\vspace{0.0cm}\hspace{0.6cm}(\textit{g})\hspace{4.2cm}(\textit{h})\hspace{4.2cm}(\textit{i})\\\vspace{-0.3cm}\\
  \includegraphics[width=0.33\linewidth]{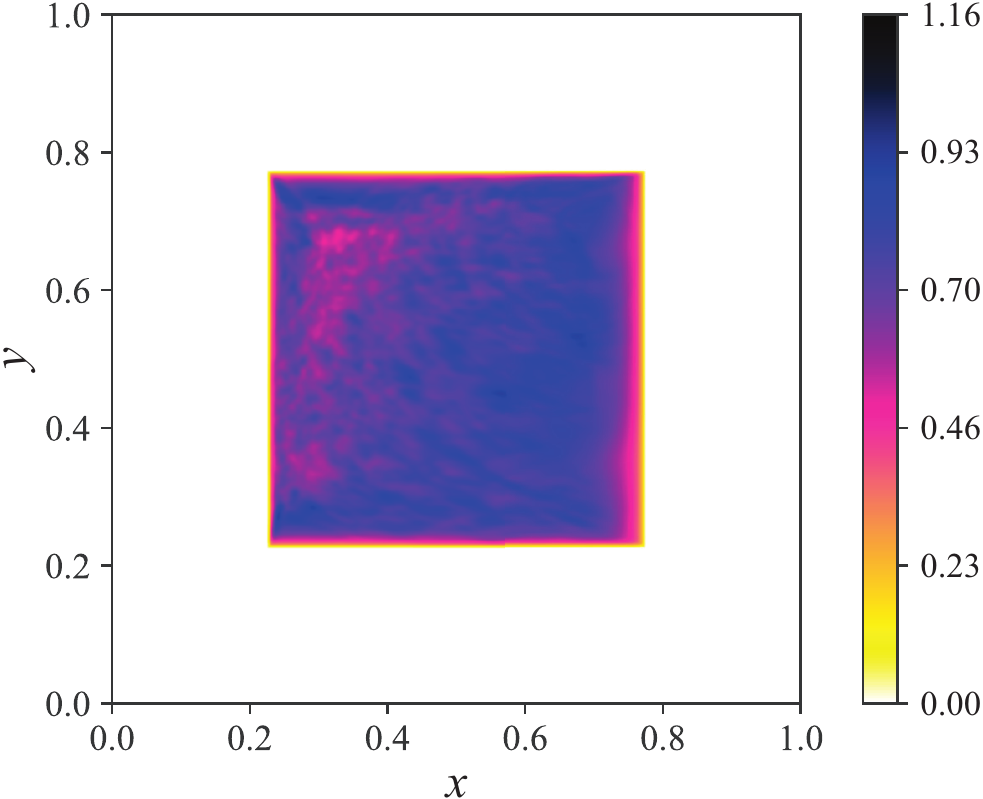}
  \includegraphics[width=0.33\linewidth]{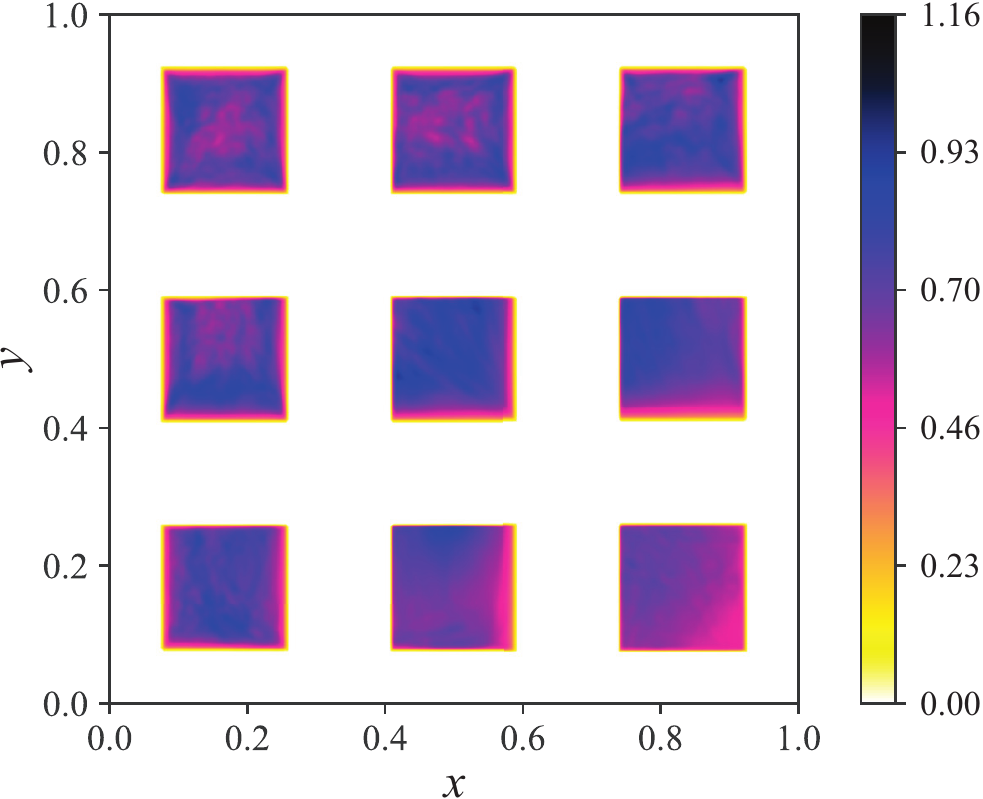}
  \includegraphics[width=0.33\linewidth]{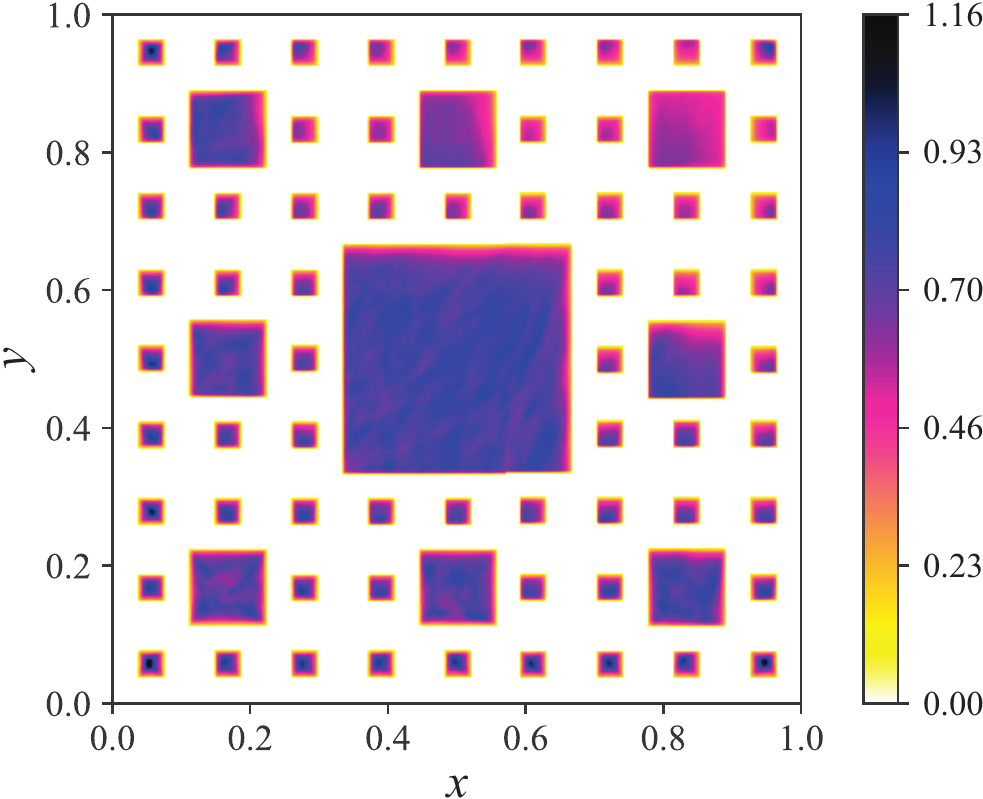}
  \caption{Time-averaged normalized thermal boundary layer thickness $\delta^{*}$ for (\textit{a}-\textit{c}) $\Ray=10^{7}$, (\textit{d}-\textit{f}) $\Ray=10^{8}$ and (\textit{g}-\textit{i}) $\Ray=1.1\times10^{9}$. Pure white corresponds to the insulating regions with zero values of local Nusselt number.}
\label{fig:kd}
\end{figure}

\begin{figure}
  \includegraphics[width=0.9\linewidth]{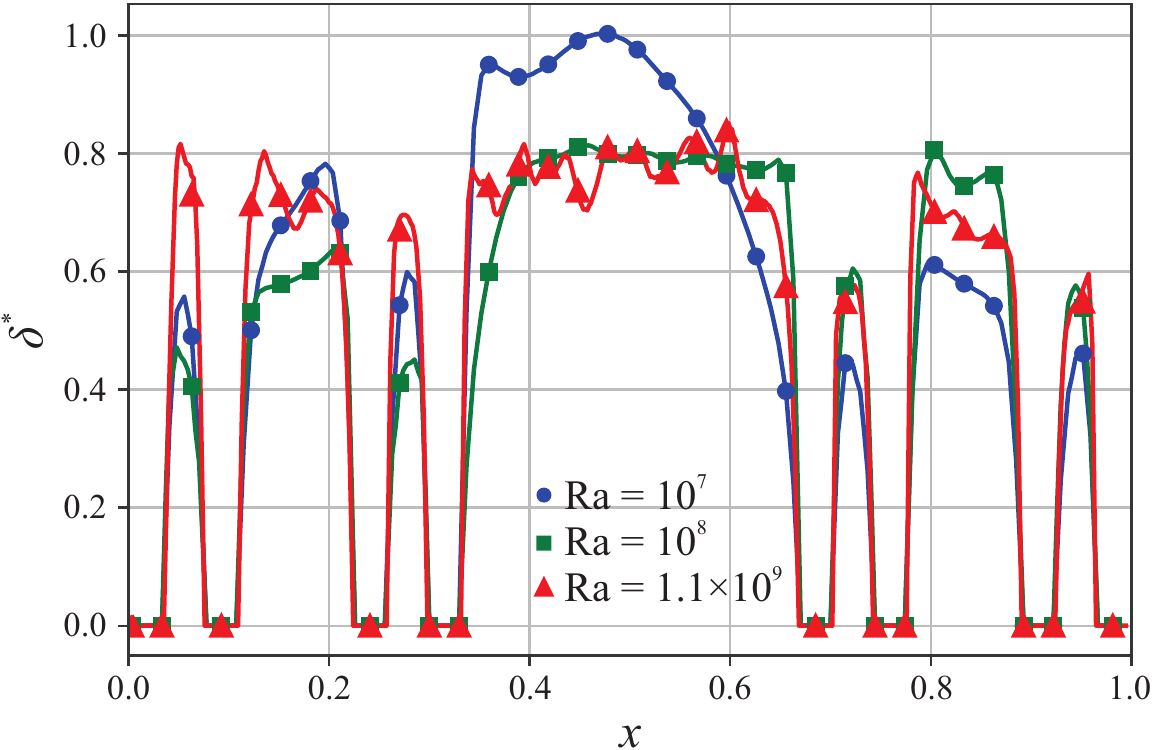}
  \caption{Normalized profiles of thermal boundary layer thickness $\delta^{*}$ for $y=0.5$ in case of fractal conducting-adiabatic pattern. }
\label{fig:profile}
\end{figure}

The thermal boundary layer is the key issue for understanding the heat transfer processes. Here, using a number of different configurations of conducting plates, including multiscale (fractal) cases we focus our attention on the structure of the boundary layer and its variation with different distribution of the conducting plates. In order to obtain spatial distribution of the thermal boundary layer thickness we used local Nusselt number defined as
\begin{equation}
  \Nuy_{l} = \left.-\dfrac{\partial \theta}{\partial z}\right|_{z=0}.
  \label{local_Nu}
\end{equation}
Then, we can estimate the thickness of the thermal boundary layer as $\delta_{\theta}=1/2\Nuy_{l}$. We need to note that adiabatic boundary condition at insulating regions results in zero values of local Nusselt number. The reference system is classical RBC case so for the further analysis we provide distributions of time-averaged normalized boundary layer thickness $\delta^{*}=\delta_{\theta}/\delta_0$, where $\delta_0$ is the mean boundary layer thickness in RBC case for the same value of Rayleigh number. The typical distributions of $\delta^{*}$ for the steady-state stage for different configurations are presented in figure~\ref{fig:kd}. Values of $\delta^{*}$ smaller than unity indicates boundary layers thinner than for RBC case. From figure~\ref{fig:kd} we see evident influence of the large-scale circulation (LSC) on the thermal boundary layer structure. It is known that descending relatively cold fluid results in a more thin boundary layer in comparison with the region of ascending hot fluid. Close inspection of the boundary layer structure reveals another important feature. There are large gradients of $\delta^{*}$ in the peripheral parts of the discrete hot plates. For better illustration we provide profiles of $\delta^{*}$ for $y=0.5$ and different values of $\Ray$ in case of fractal conducting-adiabatic pattern (see figure~\ref{fig:profile}).
\begin{figure}
\vspace{0.35cm}\hspace{0.6cm}(\textit{a})\hspace{4.1cm}(\textit{b})\hspace{4.1cm}(\textit{c})\\\vspace{-0.2cm}\\
\includegraphics[width=0.33\linewidth]{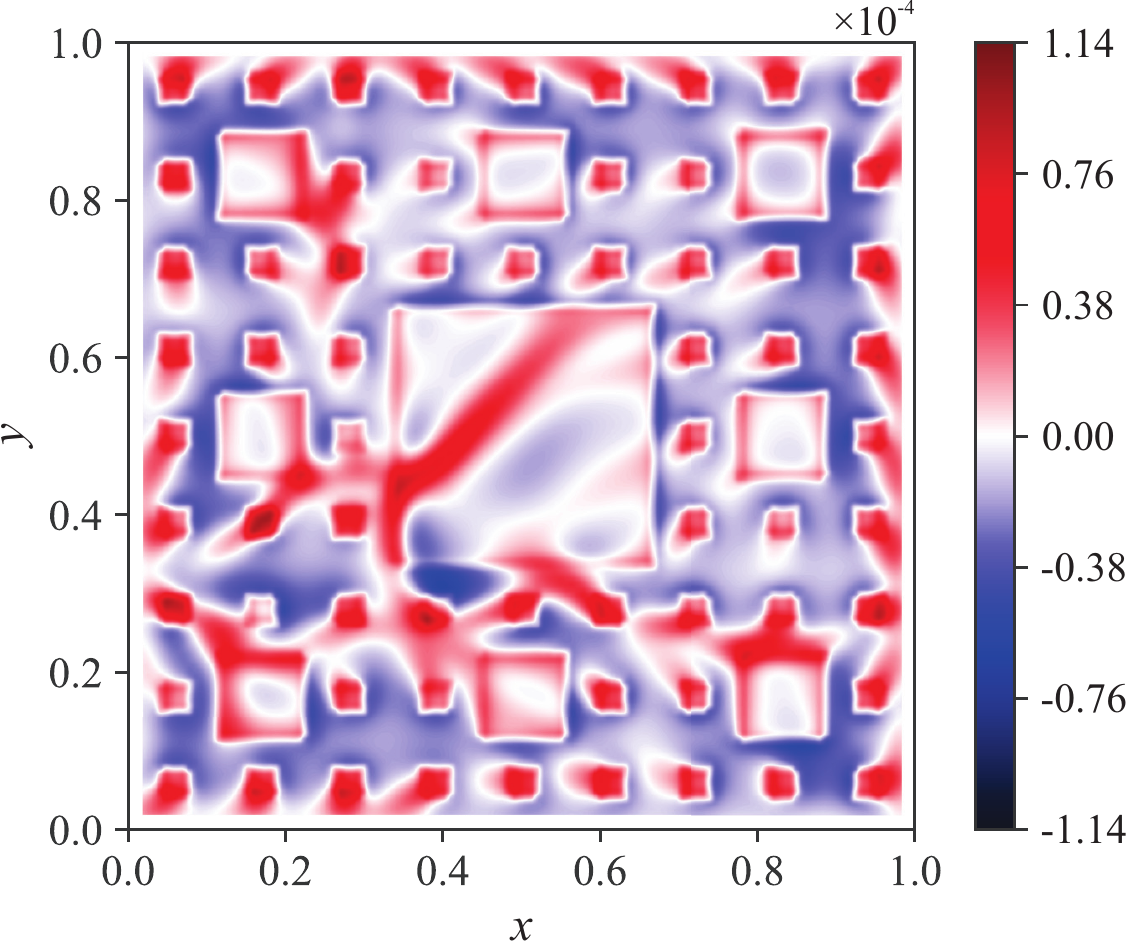}
\includegraphics[width=0.33\linewidth]{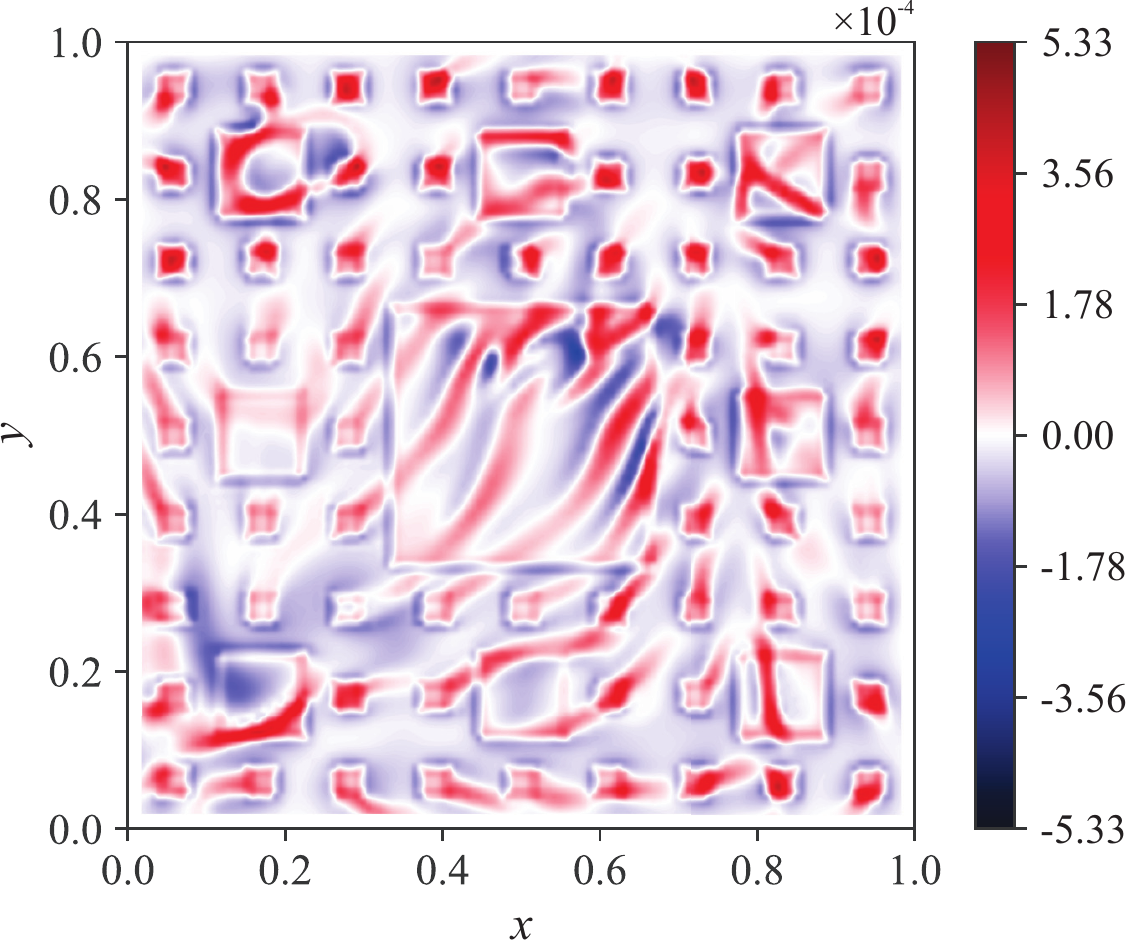}
\includegraphics[width=0.33\linewidth]{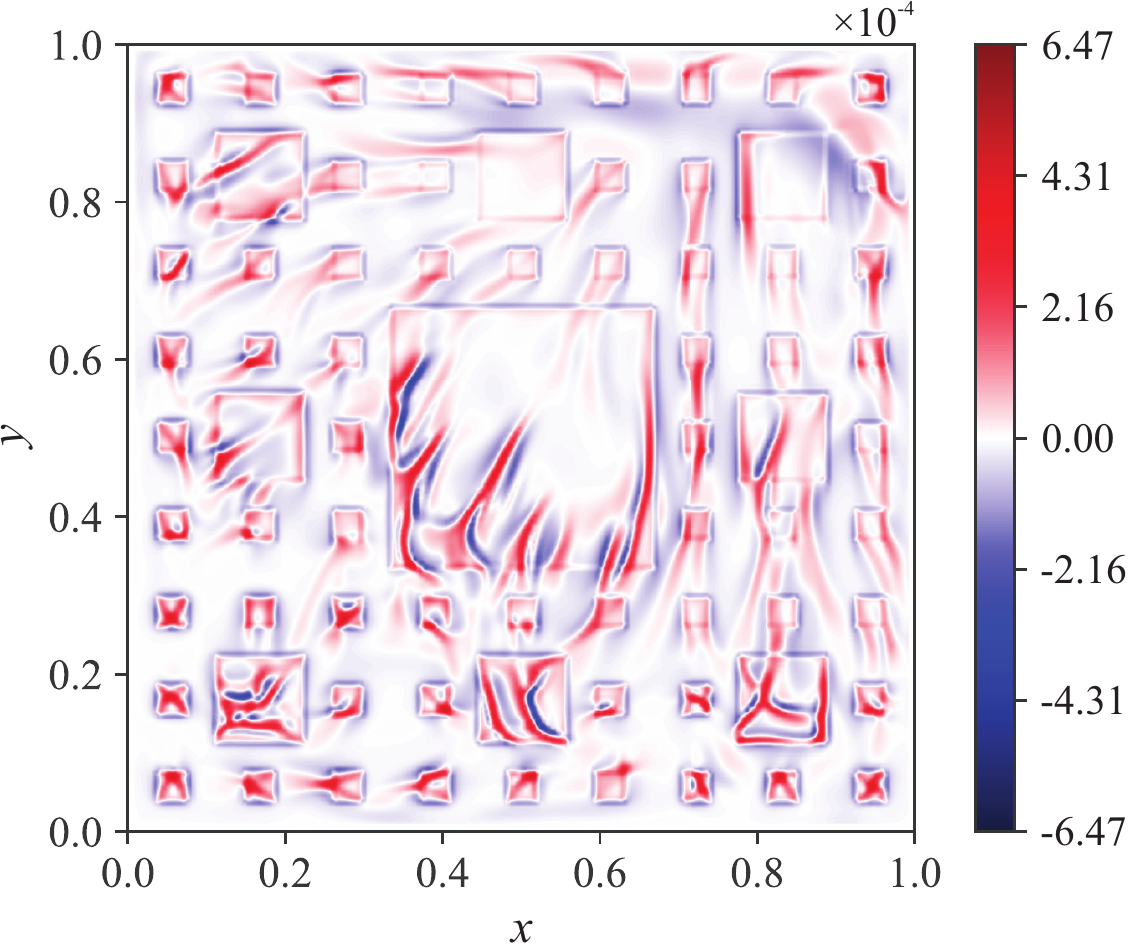}
\caption{Instantaneous fields of vertical velocity $u_{z}$ in a horizontal cross-section at: (\textit{a}) - $z=0.001$, $\Ray=10^{7}$, (\textit{b}) - $z=0.001$, $\Ray=10^{8}$ and (\textit{c}) - $z=0.0007$, $\Ray=1.1\times 10^{9}$.}
\label{fig:NonHomVel}
\end{figure} 
Substantial inhomogeneity of thermal boundary layer is a result of several factors. At first we see the imprint of applied boundary conditions -- the conducting-adiabatic pattern. As we mentioned earlier the appearance of LSC also has a strong influence on the boundary layer structure. And the last factor is a formation of small-scale motions induced by local horizontal temperature gradients from discrete hot plates. The structure of these small-scale flows embedded in the LSC is illustrated by vertical component of velocity $u_{z}$ for different $\Ray$ and for fractal configuration in figure~\ref{fig:NonHomVel}. What we see in figure~\ref{fig:kd} and figure~\ref{fig:profile} is a result of combined action of all described effects. The formation of the boundary layers over conducting plates in the considered configuration is similar to the case of forced convection over the hot plate, where the boundary layer thickness is not uniform and $\delta_{\theta}$ grows further from the border of the plate along the flow. In figure~\ref{fig:profile} we also see that the shape of $\delta^{*}$ profile strongly depends on the plate size. When it is relatively large than the boundary layer has a plateau in the central part of the plate. It is worth to mention that even when the  plateau is noticeable (the largest plate) the thickness of the boundary layer is close to the RBC case only for the small $\Ray$ and for the higher $\Ray$ its normalized value is about $\delta^{*}\approx0.8$.

The complex structure of the flow which is a superposition of different large and small-scale modes results in asymmetric boundary layer profiles. Beside asymmetry there is evident dependence of $\delta^{*}$ on the size of the hot plate. The horizontal size of the plates is a limiting factor for the thickness of the boundary layer, so the smaller is the plate the thinner is the boundary layer. This means that heat flux should increase with decreasing of the conducting plate size. Using multiscale configuration we can check this assumption for $\Ray$ from $10^{7}$ to $10^{9}$. For this purpose we provide ratios of heat fluxes over hot plates in fractal configuration and mean heat flux for corresponding RBC case (table~\ref{tab:2}). There is remarkable difference between heat fluxes in RBC case and fractal configuration. Partially this difference is a result of variation of the mean temperature of the fluid due to disbalance of heat fluxes near the top and bottom boundaries during the non-stationary stage. It deserves special attention but in the present study we focus on comparison of heat fluxes for conducting plates of different scales. The provided data clearly shows that heat transfer efficiency increases with decreasing of the size of conducting plate. For  $\Ray=10^{7}$ the heat flux of the smallest plates exceeds the one for the largest plate more than twice. Observed difference between heat fluxes from the plates of various scales decreases with increasing of $\Ray$. This result is expected because increasing of $\Ray$ provides a more thin thermal boundary layer, which increases ratio of the size of the plate and thickness of the boundary layer, so the plates become effectively larger. The same effect provides increasing the plate size for the fixed $\Ray$. 

%\begin{figure}
%\center{\includegraphics[width=0.50\linewidth]{Fig_1.eps}}
%%\caption{Sketch of the computational domain.}
%%\label{fig:Schema}
%\end{figure}
%\begin{figure}
%\vspace{0.35cm}\hspace{0.6cm}(\textit{a})\hspace{4.1cm}(\textit{b})\hspace{4.1cm}(\textit{c})\\\vspace{-0.2cm}\\
% \includegraphics[width=0.33\linewidth]{Fig_3c.eps}
%  \includegraphics[width=0.33\linewidth]{Fig_3f.eps}
%  \includegraphics[width=0.33\linewidth]{Fig_3i.eps}
%\vspace{0.35cm}\hspace{0.6cm}\\
%\includegraphics[width=0.33\linewidth]{Fig_5a.eps}
%\includegraphics[width=0.33\linewidth]{Fig_5b.eps}
%\includegraphics[width=0.33\linewidth]{Fig_5c.eps}
%%\caption{Instantaneous fields of vertical velocity $u_{z}$ in a horizontal cross-section at: (\textit{a}) - $z=0.001$, $\Ray=10^{7}$, (\textit{b}) - $z=0.001$, $\Ray=10^{8}$ and (\textit{c}) - $z=0.0007$, $\Ray=1.1\times 10^{9}$.}
%%\label{fig:NonHomVel}
%\end{figure} 

\section{Conclusions}
\label{sec:concl}

In the present paper we examine the structure of the thermal boundary layer in case of mixed boundary conditions. Recent numerical simulations \citep{RipesiJFM2014, DennisJFM2018} revealed very interesting feature of this specific case of Rayleigh-B{\'e}nard convection -- increasing of heat flux with spatial frequency of conducting-adiabatic pattern. It was assumed that this growth of heat flux (and Nusselt number) was provided by gradual spatial homogenization of the thermal boundary layer due to horizontal heat fluxes \citep{DennisJFM2018}. Alternative explanation of this phenomenon can be based on formation of strongly non-uniform thermal boundary layer which is predominantly more thin over conducting plates in comparison with a classical RBC case. Our numerical simulation including multiscale configuration support this scenario. We have shown that thickness of the thermal boundary layer strongly depends on the size of the conducting plates and can be substantially smaller than for a classical RBC. This effect increases the heat flux with decreasing the size of hot plates, which corresponds to the increasing of spatial frequency of conducting-adiabatic pattern. Proposed physical mechanism perfectly fits to the all known studies of Rayleigh-B{\'e}nard convection in case of mixed boundary conditions.   
\begin{table}
  \begin{center}
\def~{\hphantom{0}}
  \begin{tabular}{ccccc}
       $\Ray$     & $q_{mean}$/$q_{RBC}$  & $q_{1}$/$q_{RBC}$  & $q_{2}$/$q_{RBC}$  & $q_{3}$/$q_{RBC}$ \\[3pt]
       $10^{7}$             & 2.15        & 1.41     & 2.1     & 3.13\\
       $10^{8}$             & 1.93        & 1.47     & 1.91     & 2.53\\
       $1.1\times 10^{9}$   & 1.71       & 1.4     & 1.68     & 2.14\\
  \end{tabular}
  \caption{Ratios of heat fluxes over hot plates in fractal configuration and mean heat flux for corresponding RBC case. $q_{mean}$ -- heat flux averaged over hot plates of all scales, $q_{1}$ -- mean heat flux of largest plate, $q_{2}$ -- mean heat flux of middle-size plates, $q_{3}$ -- mean heat flux of small-scale plates, $q_{RBC}$ -- mean heat flux for a classical RBC. }
  \label{tab:2}
  \end{center}
\end{table}

\section{Acknowledgements}
\label{acknow}
The simulations have been done in frame of RFBR No 19-41-590004 project using the Triton supercomputer of the ICMM UB RAS, Perm, Russia. 

\bibliographystyle{jfm}
\bibliography{jfm}

\end{document}